\documentclass[flushrt,preprint]{aastex}
\tighten
\eqsecnum

\input psfig

\def\kms{\,{\rm km\, s^{-1}} }
\def\kmsm{\,{\rm km\, s^{-1}\, Mpc^{-1}} }

\begin{document}

\title{Constraints on the Inner Mass Profiles of Lensing Galaxies from 
Missing Odd Images} 
\author{David Rusin, Chung-Pei Ma} 
\affil{Department of Physics and
Astronomy, University of Pennsylvania, 209 S. 33rd St., Philadelphia, PA,
19104-6396}

\begin{abstract}

Most gravitational lens systems consist of two or four observable images.  The
absence of detectable odd images allows us to place a lower limit on the
power-law slope of the inner mass profile of lensing galaxies.  Using a sample
of six two-image radio-loud lens systems and assuming a singular power-law
surface density ($\Sigma \propto r^{-\beta}$) for the inner several kpc of the
mass distribution, we find that there is less than a 10\% probability that the
data are consistent with profile slopes $\beta < 0.80$. Furthermore,
individual mass modeling yields $\beta > 0.85$ for B0739+366 and $\beta >
0.91$ for B1030+074. Modeling central black holes as additional point masses
changes the constraints in these systems to $\beta > 0.84$ and $\beta > 0.83$,
respectively. The inner mass profiles of lensing galaxies are therefore not
much shallower than isothermal.

\end{abstract}

\keywords{galaxies: structure -- gravitational lensing}

\section{Introduction} 

Many properties of a galaxy can be learned from its mass profile.
Measurements of galaxies' mass profiles at outer radii have provided some of
the most compelling evidence for the existence of dark matter (e.g.\ Rubin,
Thonnard \& Ford 1980). The inner mass distributions also carry a wealth of
information, and can be readily investigated by strong lensing. For a typical
lens redshift of $z \simeq 0.5$ and reasonable cosmological parameters,
arcsecond-scale lens systems probe the galaxy mass profile out to a $3-6$
kpc. Lensing has already taught us much about the inner several kpc of
elliptical galaxy lenses by essentially ruling out constant mass-to-light
ratio models, based on poor fits to individual lens systems (Kochanek 1995;
Grogin \& Narayan 1996; Romanowsky \& Kochanek 1999) and an inability to
reproduce observed image separations (Maoz \& Rix 1993). Power-law mass
profiles with surface mass density $\Sigma(r) \propto r^{-\beta}$, and the
isothermal case ($\beta = 1$) in particular, provide improved models of
lensing mass distributions while remaining statistically consistent. Mass
modeling has thus far yielded direct constraints on the power-law profile in
only a few systems: QSO 0957+561 ($0.82 \leq \beta \leq 0.93$; Grogin \&
Narayan 1996), MG1131+0456 ($1.2 \leq \beta \leq 1.6$; Chen et al.\ 1995),
B1608+656 ($0.8 \leq \beta \leq 1.2$; Koopmans \& Fassnacht 1999), MG1654+1346
($0.9 \leq \beta \leq 1.1$; Kochanek 1995), and B1933+503 ($0.7 \leq \beta
\leq 1.0$; Cohn et al.\ 2000).  Unfortunately, most two-image lenses do not
provide a sufficient number of constraints to allow for a detailed
investigation of the galaxy mass profile.

This Letter investigates a method for placing a lower bound on the inner slope
of the mass profiles of lensing galaxies in two-image systems. Our technique
takes advantage of the lensing property that as the profile is made shallower
(relative to isothermal), a faint image close to the center of the lens
becomes more prominent (Narasimha et al.\ 1986; Blandford \& Kochanek 1987).
This image is absent in virtually all known lens systems, with the possible
exceptions of MG1131+0456 (Chen \& Hewitt 1993), in which the central
component may be emission from the lensing galaxy itself, and APM08279+5255
(Ibata et al.\ 1999), which may be a special class of imaging due to an
edge-on disk (Keeton \& Kochanek 1998). The paucity of detectable third images
therefore allows us to constrain the steepness of the lens mass profile. To
achieve this goal we use two complementary approaches. First, we compute the
frequencies at which detectable three-image systems are produced as a function
of the power-law index and axis ratio of the mass distribution (\S~2). We then
derive a limit on the characteristic slope of the inner mass profile by
performing a statistical analysis on a sample of six radio-loud two-image lens
systems (\S~3). Second, we use mass modeling to directly constrain the galaxy
mass profiles for two individual lens systems (\S~4): B0739+366 (Marlow et
al.\ 2001) and B1030+074 (Xanthopoulos et al.\ 1998).  The effect of central
black holes is also explored.

\section{Frequency of Detectable Triple Systems}

The inner mass profile of lensing galaxies is approximated as a singular
power-law ellipsoid (SPLE; e.g.\ Barkana 1998), with a scaled projected
surface mass density $\kappa(x_1, x_2) = (b /2) (x_1^2 + f^2
x_2^2)^{-\beta/2}$ where $b$ is the characteristic angular scale, $\beta$ is
the radial profile and $f$ is the axial ratio. For the isothermal case, $\beta
= 1$. The SPLE likely oversimplifies the overall mass distribution in galaxies
but is a reasonable model for the inner several kpc that are probed by strong
lensing.  We can gain some analytical insight from the spherical case ($f=1$).
For $\beta<1$ the model has critical curves at $x_{r} = [b (1-\beta ) /
(2-\beta)]^{1/\beta}$ (radial) and $ x_{t} = [b / (2-\beta) ]^{1/\beta}$
(tangential) on which the image magnification diverges, and a radial caustic
at $y_{r} = [ b (1-\beta) / (2-\beta) ]^{1/\beta} [\beta/(1-\beta)]$ inside of
which sources are triply-imaged. The third central image resides within the
radial critical curve. If $\beta\geq1$ this curve does not exist, so the
central image is trapped in the galaxy core and completely demagnified. The
ratio of the radial to tangential critical curves is
$x_r/x_t=(1-\beta)^{1/\beta}$, which increases as the slope $\beta$ is
decreased from isothermal.  A larger radial critical curve means that the
third image can form farther from the lens center, so it tends to be less
demagnified by the high central convergence. Consequently, additional images
of this type are more likely to be observed for shallow mass profiles.

Elliptical profiles with $\beta < 1$ also efficiently produce a second class
of three-image lens system: the naked cusp configuration (e.g.\ Kormann,
Schneider \& Bartelmann 1994). These can form when the tangential caustic
extends outside the radial caustic. Sources placed within the exposed
tangential caustic form three bright images on the same side of the lens, and
at similar distances from the lens center. The tangential caustic is
determined almost exclusively by the ellipticity of the potential (Blandford
\& Kochanek 1987), changing little as the profile is modified.  The radial
caustic, however, shrinks by $\sim 50\%$ (for the spherical case) when the
power-law index is decreased from $\beta=1$ to $0.8$. As a result, the
tangential caustic is more easily exposed for shallower mass profiles and
naked cusp systems grow in prominence, as demonstrated in Fig.\ 1. We refer to
these naked cusp systems as ``Type II'' and all other three-image systems as
``Type I.''  Near-isothermal models with reasonable axial ratios ($f \geq
0.4$) produce predominately Type I systems.

In this analysis we focus on the three-image systems produced by lenses with
the SPLE mass density.\footnote{The additional fifth image for quads tends to
be significantly more demagnified than the corresponding third image for
doubles (Wallington \& Narayan 1993). Therefore the absence of a fifth image
in quads leads to much poorer constraints on the mass profile.} To make use of
the information provided by unobserved odd images, one must first determine
the frequency at which observable three-image systems are produced by
power-law deflectors. To this end, we have performed Monte Carlo simulations
in which sources are placed randomly in the Type I and Type II caustic regions
of the SPLE for various combinations of $\beta$ ($\leq 1$) and $f$.  In each
trial the lens equation is numerically inverted to solve for all image
positions and magnifications. Our calculations make use of the rapidly
converging series solutions for the deflection angles and magnification
matrices of power-law mass distributions derived by Barkana (1998) and
implemented in the ``FASTELL'' software package.

We illustrate in Fig.~2a--2c the unbiased fraction of three-image systems with
$\mu_1/\mu_3 \leq 100$, as a function of $\beta$ and $f$. The dependence on
$\beta$ of the curves in Fig.~2 can be understood as follows.  For $\beta=1$,
the radial critical curve degenerates to a point and the third image is
infinitely demagnified, so $\mu_1/\mu_3\rightarrow \infty$, and the fraction
of Type I configurations ($p_I$) is zero. As the profile is made more shallow,
the radial critical curve grows in size, the third image inside the radial
critical curve becomes brighter, and $p_I$ increases initially as $\beta$
decreases from 1 (Fig.~2a).  As $\beta$ is decreased further, however, Fig.~2a
shows that the Type I fraction begins to decline beyond a cutoff that depends
on the axial ratio $f$.  For $\beta$ below this cutoff, Type II systems begin
to dominate (Fig.~1 and Fig.~2b), and a large fraction of three-image systems
should appear as three relatively bright images on the same side of the lens,
which has never been observed.  A profile much shallower than isothermal is
therefore highly disfavored (Fig.~2c).


\section{Statistical Constraints on Mass Profiles from Radio Data}

Radio-loud gravitational lens systems are ideal targets with which to search
for faint odd images.  Such systems can be investigated with high dynamic
range maps that are not contaminated by galaxy emission which could mask the
presence of faint additional images near the lens center.  The most extensive
radio lens search is the combined Jodrell-VLA Astrometric Survey (JVAS;
Patnaik et al.\ 1992; Browne et al.\ 1998; Wilkinson et al.\ 1998; King et
al.\ 1999) and Cosmic Lens All-Sky Survey (CLASS; e.g.\ Myers et al.\
1999). Lens candidates are selected from snapshot imaging with the Very Large
Array (VLA) and undergo deep follow-up observations using the Multi-Element
Radio-Linked Interferometer Network (MERLIN) and the Very Long Baseline Array
(VLBA), each of which offers excellent sensitivity (rms $\sim 50$
$\mu$Jy/beam). The compact nature of the JVAS/CLASS lensed sources should
allow the detection of very faint additional images. Since typical CLASS lens
systems have a primary component with $S_1 \geq 25$ mJy at 5 GHz, a
conservative 5 rms limit should ensure the detection of any faint third image
such that the magnification ratio is $\mu_1 / \mu_3 \leq 100$. A number of
radio lens systems are significantly brighter than $25$ mJy, and in such cases
demagnified central components would be more readily detectable.

The combined JVAS/CLASS lens survey currently contains seven two-image
gravitational lens systems. We remove B1127+385 (Koopmans et al.\ 1999) due to
its compound deflector, and take the remaining six lenses as our sample. Each
of the systems has been investigated with deep MERLIN and VLBA observations,
and no additional images have been found down to the 5 rms detection limit
of the maps.  For each lens system, Table~1 lists the flux density of the
brightest image in the most sensitive map and the corresponding constraint on
the magnification ratio ($r = S_1 /$ 5 rms).

The raw frequencies $p(\beta, f, r)$ at which detectable three-image lens
systems with $\mu_1 / \mu_3 \leq r$ are produced by the SPLE can be
calculated, as demonstrated in Fig.~2a--2c. However, these frequencies ignore
magnification bias (e.g.\ Maoz \& Rix 1993), through which lens systems with
high {\em total} magnifications preferentially appear in flux-limited samples
such as JVAS/CLASS. Highly magnified doubles tend to form when the source is
well aligned with the lens (e.g.\ Blandford \& Kochanek 1987), a situation
that also keeps the third image close to the lens center where it can be
strongly demagnified. Therefore magnification bias may dilute the probability
that lens systems in the JVAS/CLASS sample will have detectable central
images.  We account for this as follows. For a source population described by
the differential number-flux relation $N(S) \propto S^{-\eta}$, the biased
lensing cross-section for creating a system with $\mu_1 / \mu_3 \leq r$ is
given by an integral over the area enclosed by the caustics: $\sigma_B (r) =
\int \mu^{\eta - 1}(y_1, y_2) R(y_1, y_2, r) dy_1 dy_2$ where $\mu = \sum
\mu_i$ and $R$ = 1 if $\mu_1 / \mu_3 \leq r$ (0 otherwise). The biased
fraction of systems with $\mu_1 / \mu_3 \leq r$ is then simply $p_B(r) =
\sigma_B(r) / \sigma_B(\infty)$. These fractions are plotted for the SPLE as a
function of $\beta$ and $f$ in Fig.~2d--2f for $r = 100$, assuming $\eta =
2.1$ for the JVAS/CLASS radio source sample (Rusin \& Tegmark 2001). As
expected, magnification bias decreases the frequency of Type I systems, but
not overwhelmingly; e.g.\ $p_B(r=100) = 0.16$ for $f=0.7$ and $\beta=0.8$,
compared to $p(r=100) = 0.20$. On the other hand, Type II systems become more
prominent due to their high magnifications (Fig.~2e).

The probability that a lens system will have no third image down to the
detection limit $r_i$ is $1 - p_B(\beta, f, r_i)$, where $p_B = p_{BI} +
p_{BII}$. To constrain the typical inner profile slope of lensing galaxies, we
evaluate the probability $P(\beta, f) = \prod_{i=1}^{6} [1 - p_B(\beta, f,
r_i)]$ of the data given the model. We assume a moderate mass axial ratio of
$f=0.7$, which is consistent with the published models of each lens system in
the sample to $\sim 10\%$. The resulting probability is plotted in Fig.~3 as a
function of $\beta$. Note that $P=1$ for $\beta \geq 1$ because no Type I
systems are formed. As $\beta$ is decreased from isothermal the probability
falls rapidly, so small values of $\beta$ are strongly ruled out. For example,
$P < 0.10$ if $\beta < 0.8$ and $P < 0.05$ if $\beta < 0.78$. We take the
former as our statistical lower limit on the profile slope: $\beta > 0.80$.

\section{Mass Modeling of Two Lens Systems}

We now expand on our statistical result by directly constraining the profiles
of specific galaxies.  For simplicity we consider only two-image systems
lensed by an isolated elliptical galaxy whose centroid has been well
determined, i.e., B0739+366 (Marlow et al.\ 2001) and B1030+074 (Xanthopoulos
et al.\ 1998). Modeling of these systems is performed with data available in
the literature. The position of the lens galaxy is fixed according to the
optical imaging. In each trial we set the power-law index $\beta$ to a test
value and optimize the remaining five free model parameters: the galaxy
normalization, axial ratio, position angle, and the source coordinates.  Since
the radio data provide five constraints (two sets of image coordinates and a
flux density ratio), the best-fit model can reproduce the image properties
exactly. At the end of each trial the recovered source position is numerically
inverted to solve for all images. The magnification of the third (unseen)
image relative to the brightest image is then compared to current detection
limits, as listed in Table 1.  The predicted ratios $\mu_1 / \mu_3$ as a
function of mass profile are plotted in Fig.~4. The lack of detectable third
images implies that $\beta > 0.85$ for B0739+366 and $\beta > 0.91$ for
B1030+074.

There is significant evidence that black holes exist at the centers of
elliptical galaxies.  Mao, Witt \& Koopmans (2001) demonstrate that if lensing
galaxies are modeled as an isothermal ellipsoid with a finite core radius, the
addition of a central black hole can steepen the inner profile and suppress
the magnification of the third image.  In our analysis, we have instead used a
singular ellipsoid with an arbitrary power-law index $\beta$ to approximate
the inner mass distribution from all forms of matter.  The above constraints
are therefore on the total contributions from baryons, dark matter, and
perhaps black holes.  However, to quantify the separate contribution from
potential black holes, we also model these two lenses by adding a central
point mass to the SPLE according to the empirical relation $M_{BH} \sim 1.3
\times 10^8 M_{\odot} (\sigma_v/200\kms)^{4.72}$ between black hole mass and
velocity dispersion (Merritt \& Ferrarese 2001), which combines the
measurements of Ferrarese \& Merritt (2000) and Gebhardt et al.\ (2000).  To
obtain the black hole mass we use the isothermal velocity dispersion required
to produce the observed image separation (which should be a good approximation
so long as we are investigating mass profiles close to isothermal), and find
$M_{BH} \sim 1.7 \times 10^7 M_{\odot}$ for B0739+366 and $M_{BH} \sim 3.7
\times 10^8 M_{\odot}$ for B1030+074. We assume a flat $\Omega_m = 0.3$
universe with $H_o = 65\kmsm$. The resulting magnification ratio curves for
each lens are plotted in Fig.~4.  We find that when central black holes are
modeled separately as point masses, the constraints become $\beta > 0.84$ for
B0739+366 and $\beta > 0.83$ for B1030+074, a less than 10\% change from the
pure SPLE case. Using the shallower $M_{BH}-\sigma_v$ relation from Gebhardt
et al.\ (2000) we find nearly identical results: $\beta > 0.83$ for B0739+366
and $\beta > 0.84$ for B1030+074. Note that the weaker profile constraints in
models with a black hole are largely due to the complete demagnification of
the central image for steeper profiles (Fig.~4), an effect similar to that
described in Mao et al.\ (2001).

\section{Discussion}

Using the detection limits for faint third images in six radio-loud
doubly-lensed systems from JVAS/CLASS, and assuming an SPLE surface mass
density ($\Sigma\propto r^{-\beta}$), we have obtained a lower bound on the
typical power-law index $\beta$ for the inner mass profiles of lensing
galaxies, which are predominately large ellipticals.  For $\beta < 0.80$ the
probability of the data is $P < 0.10$, so we take $\beta = 0.80$ to be the
statistical lower limit on the characteristic profile slope. We have also used
mass modeling and the lack of a detectable third image to constrain the
profiles of two lensing galaxies that have well determined centroids.  We find
$\beta > 0.85$ for B0739+366 and $\beta > 0.91$ for B1030+074 when a singular
ellipsoid is used to approximate the lensing galaxy.  If the lensing galaxy is
assumed to harbor a central black hole which is modeled separately as a point
mass, we find only slightly different bounds of $\beta > 0.84$ for B0739+366
and $\beta > 0.83$ for B1030+074. These results argue that the total mass in
the form of baryons, dark matter, and perhaps black holes at the inner parts
of lensing galaxies follows a profile that is not much shallower than
isothermal.

Limits on the profile slope could be improved by increasing the number of lens
systems in the analysis, and in the future we may expand our sample to include
new lenses discovered in the PMN survey (e.g.\ Winn et al.\ 2000). Little
improvement, however, is expected by obtaining deeper maps of existing
lenses. Our Monte Carlo simulations demonstrate that detectable three-image
systems with small $\mu_1/\mu_3$ are much more common than those with large
$\mu_1/\mu_3$.  As a result, deeper observations quickly lead to diminishing
returns for the statistical profile constraints. In addition, near-isothermal
profiles ($0.95 < \beta < 1$) produce extremely faint third images that would
be difficult to detect with any instrument.

In conclusion, the lack of detectable odd images rules out a large region of
shallow profiles for the inner several kpc of galaxy mass distributions, and
therefore provides corroborating evidence for the popular isothermal model
favored by stellar dynamics (Rix et al.\ 1997), studies of elliptical galaxies
with X-ray halos (Fabbiano 1989), and previous constraints from gravitational
lensing. We stress that our result applies only to the total profile, as
lensing cannot distinguish between the luminous and dark mass constituents.
It is interesting to compare our constraints with the light profiles of local
ellipticals observed using the Hubble Space Telescope (Lauer et al.\ 1995).
The latter tend to be quite shallow ($\beta \la 0.5$) at small radii and fall
off sharply (with $\beta \ga 1.3$) beyond a break radius of $\la 300$ pc (Byun
et al.\ 1996), which would be only a few tens of milliarcseconds at a typical
lens redshift of $z \simeq 0.5$.  It should therefore be noted that what is
considered the {\it inner} region of a few arcseconds for strong lensing
corresponds to the {\it outer}, steeper part of the Byun et al.\ profile.  If
the dark matter profiles indeed follow $\beta\la 0.5$ as reported in various
numerical simulations (e.g. Navarro, Frenk \& White 1997; Moore et al. 1999),
our results then indicate that the shallower dark matter profiles conspire
with the steeper optical profiles to produce a near-isothermal mass
distribution in the inner few kpc of lensing galaxies.

\acknowledgements

This paper would not have been possible without the hard work of everyone in
the JVAS/CLASS team. We thank Paul Schechter, Scott Tremaine, Jim Fry and
Michael Fall for interesting discussions, and the referee for useful
comments. D. R. acknowledges support from the Zaccheus Daniel Foundation.
C.-P. M. acknowledges support of an Alfred P. Sloan Foundation Fellowship, a
Cottrell Scholars Award from the Research Corporation, a Penn Research
Foundation Award, and NSF grant AST 9973461.

\clearpage
\bigskip
\bigskip
\bigskip
\begin{table*}
\begin{tabular}{ l c c c c c l }
\hline \hline 
Lens & Instrument & $S_1$ (mJy) & rms ($\mu$Jy/beam) &
$r$ & Reference\\
\hline
B0218+357 & MERLIN 5 GHz  & 842 & 82 & 2054 & Biggs et al.\ 1999\\
B0739+366 & MERLIN 5 GHz  &  27 & 45 &  120 & Marlow et al.\ 2001\\
B1030+074 & MERLIN 5 GHz  & 364 & 61 & 1193 & Xanthopoulos et al.\ 1998\\
B1152+199 & MERLIN 5 GHz  &  53 & 72 &  147 & Myers et al.\ 1999\\
B1600+434 & MERLIN 5 GHz  &  44 & 61 &  144 & Koopmans, de Bruyn \& Jackson 1998\\
B2319+051 & VLBA 1.7 GHz  &  67 & 54 &  248 & Rusin et al.\ 2001\\
\hline
\end{tabular}
\caption{The radio lens sample. Flux densities for the brightest images are
from the highest dynamic range maps available for each lens. If this image
contains multiple subcomponents (e.g.\ B2319+051), we take $S_1$ to be the sum
of the subcomponent flux densities, as a faint central component would likely
be unresolved. More information about the MERLIN 5 GHz observations is
presented in Norbury et al.\ 2001. The magnification ratio detection limit $r$
is defined as $S_1 / 5$ rms.}
\end{table*}

\clearpage
\begin{figure*}
\begin{tabular}{cc}
\psfig{file=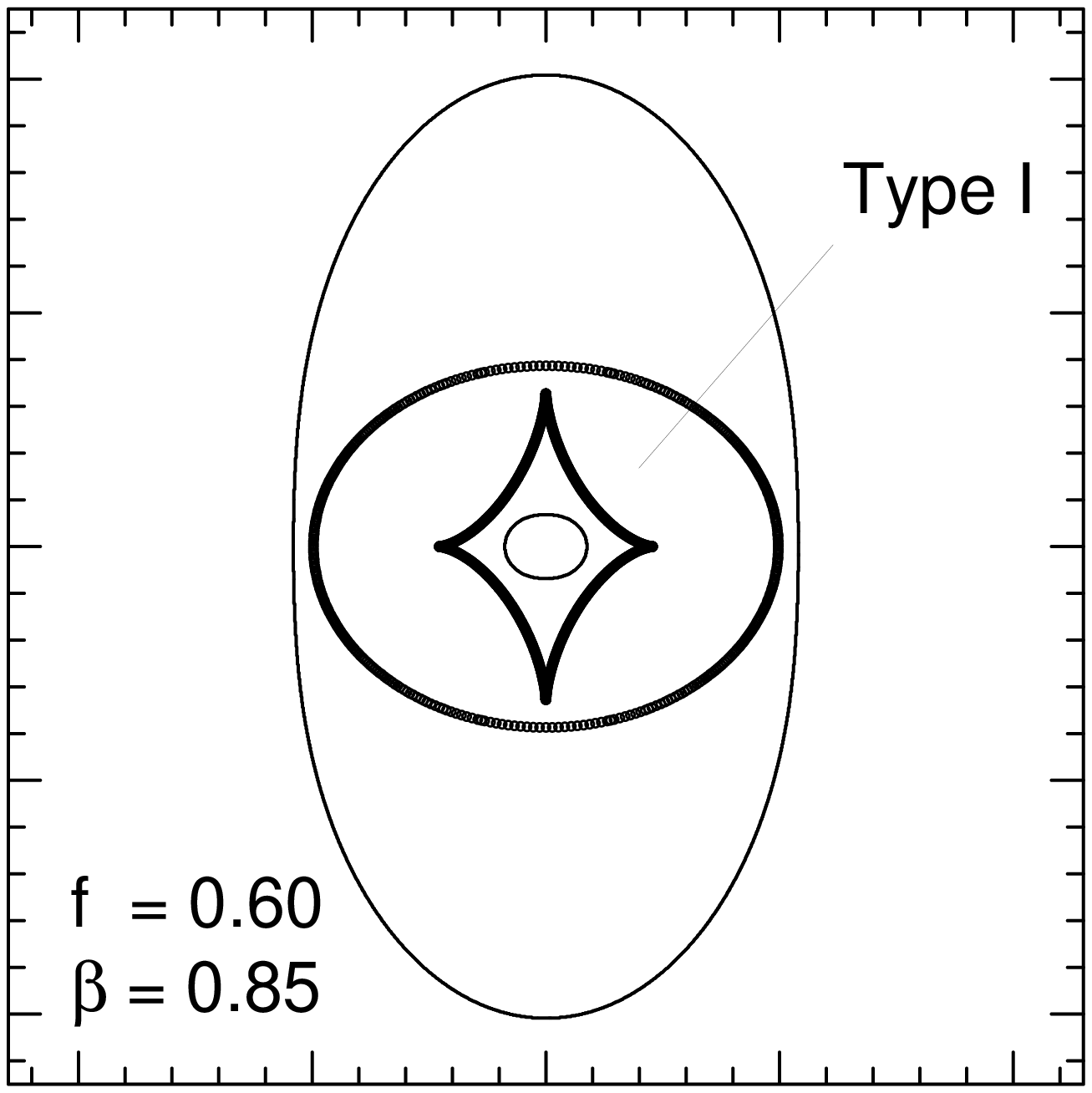,width=2in} &
\psfig{file=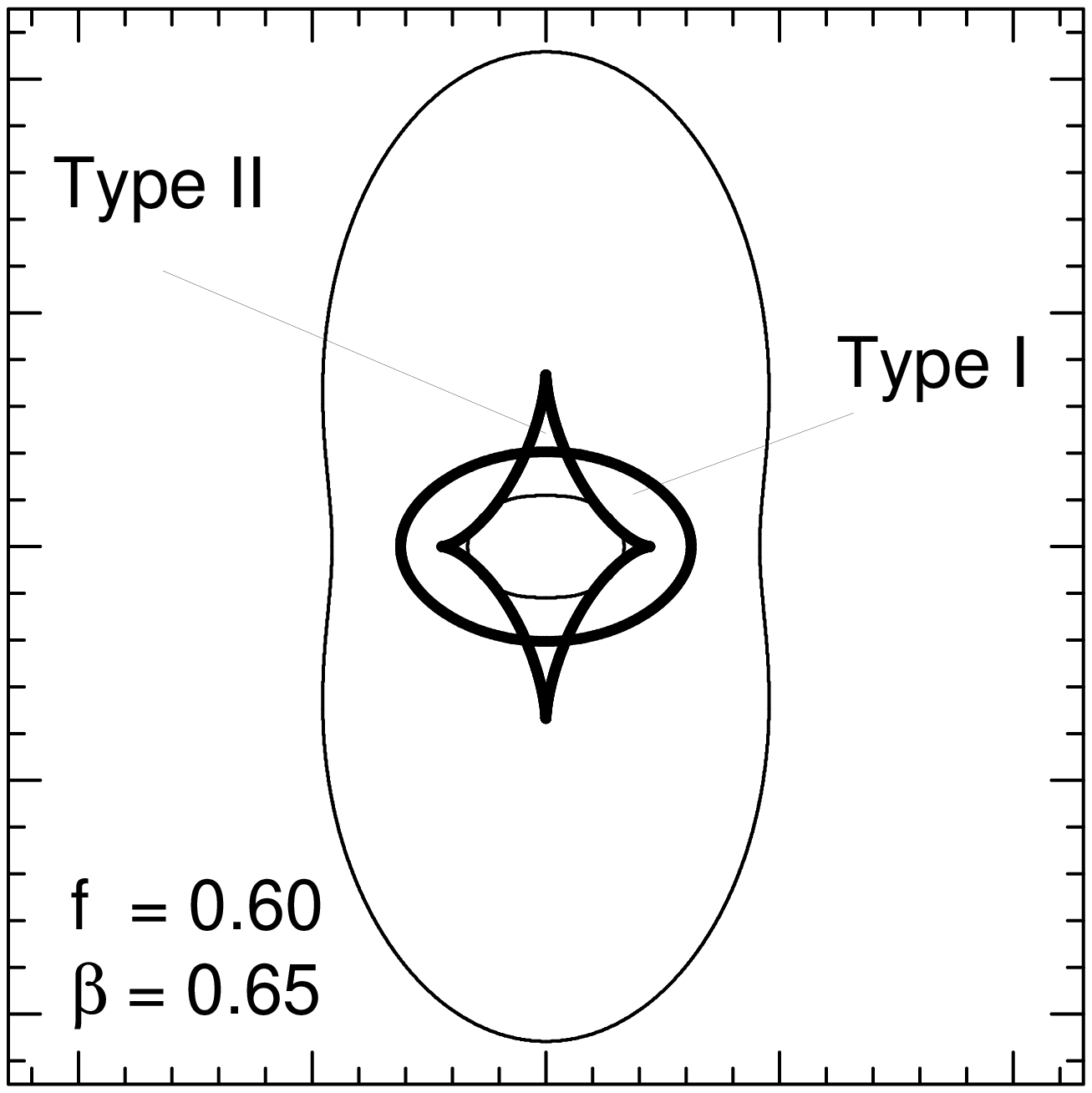,width=2in}\\
\end{tabular}
\figurenum{1}
\caption{Lensing properties of the singular power-law ellipsoid (for axis
ratio $f = 0.6$). Caustics (dark lines) separate different imaging
regions. The critical curves are drawn with thin lines. Note how decreasing
$\beta$ increases the size of the radial (inner) critical curve, which leads
to more Type I systems. At the same time, the size of the radial (elliptical)
caustic is decreased while the tangential (astroid) caustic remains unchanged,
creating Type II systems.}
\end{figure*}

\clearpage
\begin{figure*}
\begin{tabular}{c}
\psfig{file=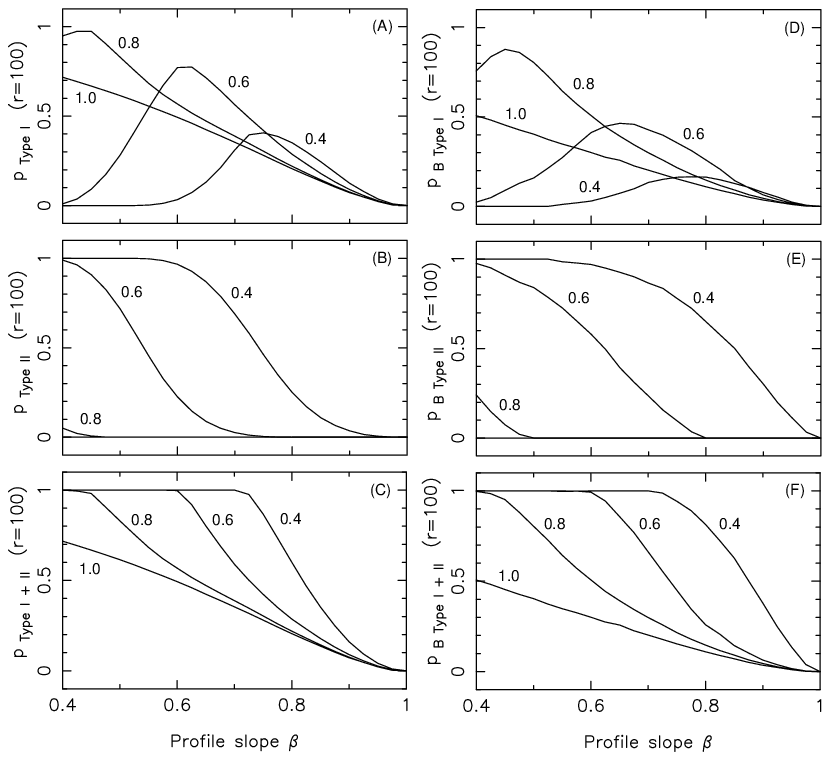,width=5.5in}\\
\end{tabular}
\figurenum{2}
\caption{Fraction of three-image configurations with $\mu_1 / \mu_3 \leq 100$,
as a function of the power-law index $\beta$ for the inner mass
profile. Different axial ratios are marked. Left: Unbiased frequencies for
Type I (a), Type II (b) and Type I + Type II (c) systems. Right:
Magnification-biased frequencies for Type I (d), Type II (e) and Type I + Type
II (f) systems.}
\end{figure*}

\clearpage
\begin{figure*}
\psfig{file=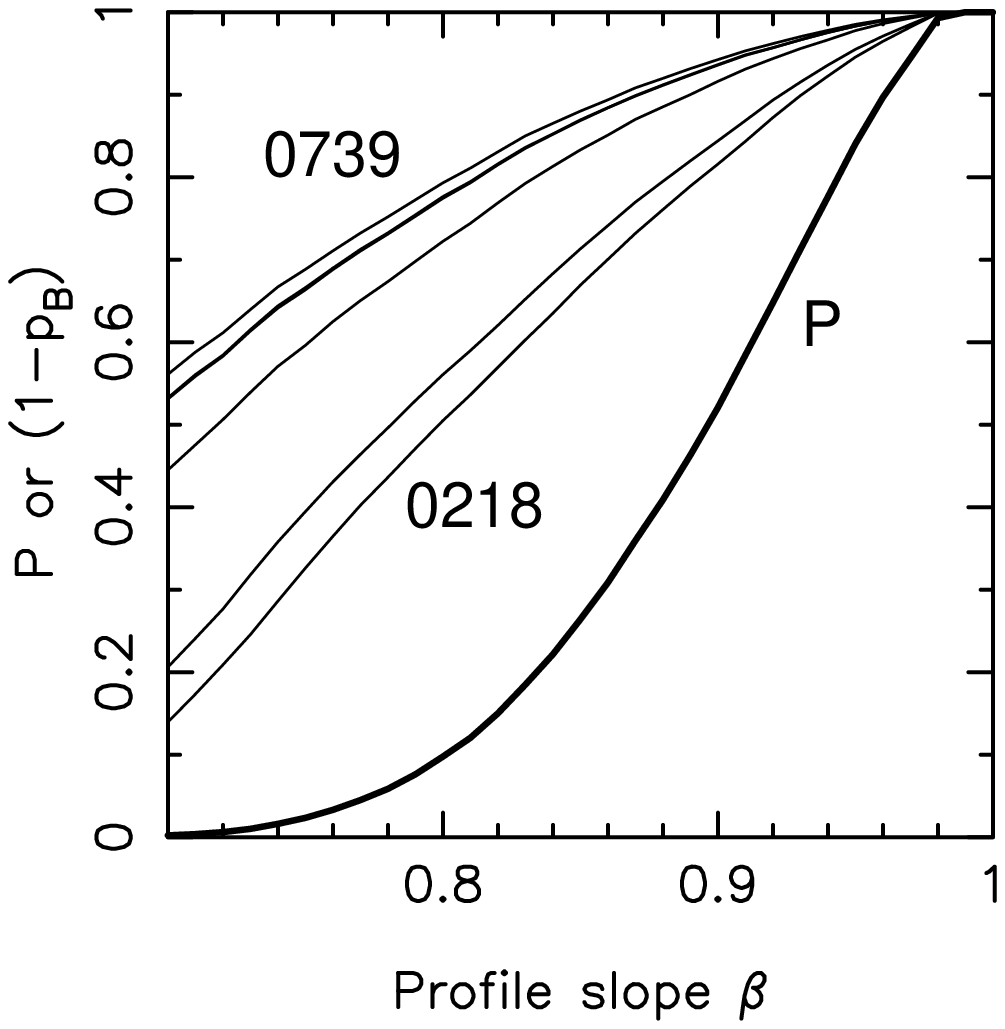,width=2.6in}
\figurenum{3}
\caption{Statistical constraints on the profile slope. Thick line represents
the probability ($P$) of the radio data as a function of $\beta$. Thin lines
represent the contributions from each of the six lenses ($1-p_B$), top to
bottom in order of increasing r from B0739+366 to B0218+357. We find $P <
0.10$ if $\beta < 0.80$ and $P < 0.05$ if $\beta < 0.78$, so shallow inner
profiles are strongly excluded by the data.}
\end{figure*}

\begin{figure*}
\psfig{file=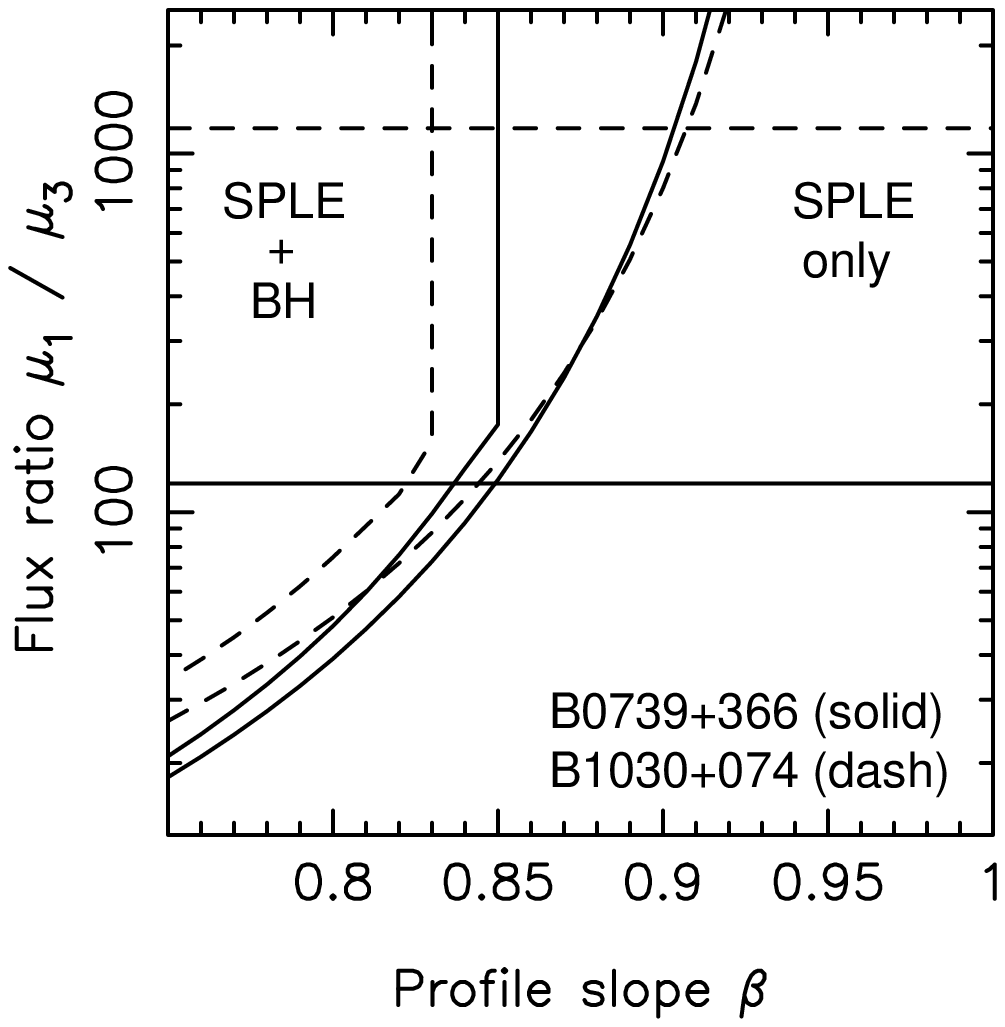,width=2.6in}
\figurenum{4}
\caption{Flux ratio $\mu_1 / \mu_3$ as a function of $\beta$ from the mass
modeling of B0739+366 (solid lines) and B1030+074 (dash lines). Leftmost
curves denote models that include a central black hole, according to the
relation of Merritt \& Ferrarese (2001). Rightmost curves are for a pure
SPLE. Note that a black hole destroys the third image for $\beta > 0.85$ in
B0739+366 and $\beta > 0.83$ in B1030+074. The respective detection limits for
third images are marked as horizontal lines and determine the constraints on
$\beta$.}
\end{figure*}

\end{document}